\documentclass[aps,
               prb,
               twocolumn,
               10pt,
               longbibliography,
               floatfix,
               nofootinbib,
               superscriptaddress]{revtex4-2}
\usepackage{bm}
\usepackage{color}
\usepackage{xcolor}
\usepackage{hyperref}
\hypersetup{
    colorlinks,%
    citecolor=blue,%
    filecolor=blue,%
    linkcolor=blue,%
    urlcolor=blue
}

\usepackage[normalem]{ulem}
\usepackage{amsfonts}
\usepackage{amsmath}
\usepackage{mathtools}
\usepackage{physics}
\usepackage{subfigure}

\begin{document}

\title{Tailoring the van der Waals interaction with rotation}

\author{H. S. G. Amaral}
\email{helenaamaral@pos.if.ufrj.br}
\affiliation{Instituto de F\'{i}sica, Universidade Federal Fluminense, 24210-346, Niter\'oi, Rio de Janeiro, Brazil}
\affiliation{Instituto de F\'{i}sica, Universidade Federal do Rio de Janeiro, 21941-972, Rio de Janeiro, Rio de Janeiro,  Brazil}
\author{P. P. Abrantes}
\affiliation{Instituto de F\'{i}sica, Universidade Federal Fluminense, 24210-346, Niter\'oi, Rio de Janeiro, Brazil}
\affiliation{Instituto de F\'{i}sica, Universidade Federal do Rio de Janeiro, 21941-972, Rio de Janeiro, Rio de Janeiro,  Brazil}
\author{F. Impens}
\affiliation{Instituto de F\'{i}sica, Universidade Federal do Rio de Janeiro, 21941-972, Rio de Janeiro, Rio de Janeiro,  Brazil}
\author{P. A. Maia Neto}
\affiliation{Instituto de F\'{i}sica, Universidade Federal do Rio de Janeiro, 21941-972, Rio de Janeiro, Rio de Janeiro,  Brazil}
\author{R. de Melo e Souza}
\email{reinaldos@id.uff.br}
\affiliation{Instituto de F\'{i}sica, Universidade Federal Fluminense, 24210-346, Niter\'oi, Rio de Janeiro, Brazil}

\begin{abstract}
We report a systematic procedure to engineer the van der Waals force between levitated nanoparticles in high vacuum by setting them into a fast rotation.  By tuning the rotation frequency close to a polaritonic resonance, we can significantly enhance the van der Waals attraction. In addition, for frequencies slightly beyond resonance, 
rotation can 
change the nature of the interaction from attraction to repulsion. Rotational Doppler shifts effectively modify the frequency-dependent polarizability of the nanoparticles, thereby reshaping their mutual interaction. As a concrete and realistic example, we consider spinning barium strontium titanate nanoparticles at state-of-the-art rotation frequencies and demonstrate a modification of the force within the sensitivity of current experimental techniques.
\end{abstract}

\maketitle

\textit{Introduction}---The progress in the field of levitodynamics has opened the way to probing weak interactions between optically trapped nanospheres in high vacuum~\cite{Gonzalez-Ballestero2021, Ranjit2015, Rieser2022}. Using similar experimental platforms, the trapping laser beam can drive the nanoparticle to spin~\cite{Monteiro2018, Reimann2018,Ahn2018} at angular frequencies as high as $5$ GHz~\cite{Ahn2020}. The combination of such experimental achievements points to the possibility of tailoring the interaction between the particles by driving them into fast rotation. In this letter, we model the van der Waals (vdW) interaction between spinning nanospheres and show a significant enhancement, along with a transition from attraction to repulsion depending on the rotation frequencies and material properties. Spinning thus provides a general approach to tune at-will vdW forces between nanoparticles, applicable across diverse geometries.

VdW interactions arise from quantum fluctuations in the charge and current distributions of interacting bodies~\cite{Woods2016, Laliotis2021}. As the distance between the interacting bodies increases, electrodynamic retardation needs to be taken into account.
In this context, the vdW force is also known as the Casimir force~\cite{Genet2004, Milton2004, Galina2009, Gong2021}, as quantum vacuum fluctuations of the electromagnetic field~\cite{Milonni2013} play a major role. These ubiquitous forces play a crucial role in various fields, including atomic, molecular, and condensed matter physics, as well as engineering, chemistry, and biology~\cite{BuhmannBook}. While the impact of acceleration on quantum vacuum phenomena (e.g., the dynamical Casimir and Unruh effects) has been thoroughly discussed~\cite{Dalvit2011, Dodonov2020, Woods2021, Impens2022}, research on spinning systems remains in its early stages \cite{Sagnac2021, Manjavacas2010, Pendry2012, Sanders2019, Manjavacas2023}. 

Here, we show that the spinning of nanoparticles induces rotation-dependent terms in the quantum correlations of the interacting dipoles, which simultaneously affect their individual response and fluctuation spectrum. This significantly shifts the materials' internal resonances, drastically altering the vdW interaction between the spinning bodies. Control of these forces is known to be difficult due to their broad-band nature.
Driving the interacting bodies out of thermal equilibrium provides an interesting route to engineer the vdW or Casimir force. Several non-equilibrium configurations have been proposed~\cite{Antezza2008,Messina2011,Bimonte2011,Bimonte2015, Chen2016, Spreng2024} but only one has been experimentally demonstrated so far~\cite{Obrecht2007}.  Rotation is a particularly interesting way of driving the system out of equilibrium as it shifts the entire spectrum, which explains why it is so powerful. To address this phenomenon, we develop a fully consistent theory of quantum correlations for spinning, fluctuating dipoles, without assuming thermal equilibrium. In fact, applying the fluctuation-dissipation theorem (FDT) to rotating nanospheres leads to serious inconsistencies. This indicates that the interaction between spinning particles is inherently a nonequilibrium effect. We handle the rotation effect by employing appropriate unitary transformations, akin to those used in the study of rotating Bose-Einstein condensates (BECs)~\cite{Dalibard05, RMPFetter}.

\textit{Quantum fluctuations for a single spinning nanosphere}---In the dipole approximation, the coupling of a neutral nanosphere with the electromagnetic field is described by the polarizability tensor $\alpha_{jm}(t-t')= (i/\hbar)\theta(t-t')\left \langle \left[ d_j(t),d_m(t') \right]\right\rangle$ and the Hadamard Green's function $\eta_{jm}(t-t') = (1/\hbar) \left\langle \left\{d_j(t),d_m(t')\right\} \right\rangle$. They capture the linear response and the fluctuations of the nanosphere's dipole ${\bm d}$, respectively. A key ingredient of our formalism is understanding how the nano-sphere spinning modifies these correlation functions. Previous works~\cite{Manjavacas2010, Pendry2012, Sanders2019, Sagnac2021, Manjavacas2023} relied on geometric considerations to obtain the polarizability tensors, assuming that the particle's quantum response is unaffected by rotation. In contrast, we adopt a more comprehensive quantum treatment in the rotating frame~\cite{Mashhoon88}, accounting for non-inertial effects such as centrifugal and Coriolis forces. This method, which has successfully explained vortex formation in BECs~\cite{RMPFetter, Dalibard05}, not only recovers the polarizabilities from Refs.~\cite{Manjavacas2010, Pendry2012, Sanders2019, Sagnac2021, Manjavacas2023} but, more importantly, provides a framework to derive a full nonequilibrium result for the Hadamard function without relying on FDT.

For a nanosphere at rest, the dipole dynamics are governed by a Hamiltonian $H^{(0)}$ that exhibits spherical symmetry, i.e., $[H^{(0)},\boldsymbol{L}]=0$ with $\boldsymbol{L}$ denoting the angular momentum operator. The dipole correlation functions for a nanoparticle at rest read $C_{ij}^{(0)}(\tau) = \langle d_i(\tau) d_j(0)  \rangle$, with the dipole operator $d_i(\tau)$ evolving under the Hamiltonian $H^{(0)}$ in the Heisenberg picture.  For a spinning particle with constant angular velocity $\mathbf{\Omega}$, the Hamiltonian in the lab frame $H^{(\boldsymbol{\Omega})}(t)$ is related to the rest frame Hamiltonian $H^{(0)}(t)$ by
 \cite{ Dalibard05, RMPFetter} $H^{(\boldsymbol{\Omega})}(t)=  H^{(0)}(t) + \boldsymbol{\Omega} \cdot \boldsymbol{L}$  (from the nanoparticle's perspective the lab frame spins at $-\mathbf{\Omega}$). In particular, the angular momentum contribution induces sidebands in the dipole correlation functions. This effect is reminiscent of rotational Doppler shifts~\cite{Power95, Birula97}, recently observed in single-ion spectroscopy~\cite{Schmiegelow24}. Once the dipole correlations $C_{ij}^{(\Omega)}(\tau)$ are obtained, the retarded and Hadamard Green functions for the dipole can be readily evaluated without invoking the FDT theorem in the transformed frame~\cite{SM}.

From now on, we focus on a spinning nanosphere, which presents isotropic polarizability and Hadamard tensors in the rest frame: $\xi_{ij}^{(0)}(\omega)=\delta_{ij} \xi(\omega)$, with $\xi$ denoting either the polarizability or the Hadamard function. Assuming the nanosphere spins around the $z-$axis with constant angular frequency $\boldsymbol{\Omega}=\Omega\boldsymbol{\hat{z}}$, the tensors in the lab frame and in the frequency domain are given by~\cite{SM}
\begin{align}
    \xi_{xx}^{(\boldsymbol{\Omega})}(\omega) &= \xi_{yy}^{(\boldsymbol{\Omega})}(\omega) = \frac{\xi(\omega_{+}) + \xi(\omega_{-})}{2} \,, \label{effpol1} \\
    \xi_{xy}^{(\boldsymbol{\Omega})}(\omega) & =  -\xi_{yx}^{(\boldsymbol{\Omega})}(\omega) = \frac{i\left[\xi(\omega_{+}) - \xi(\omega_{-})\right]}{2},  \label{effpol2}
\end{align}
where $\omega_{\pm}=\omega\pm \Omega$ are the Doppler shifted frequencies and the remaining components follow by symmetry: $\xi_{kz}^{(\boldsymbol{\Omega})}(\omega)= \xi_{zk}^{(\boldsymbol{\Omega})}(\omega) = \delta_{kz} \xi_{zz}(\omega)$, for $k=x,y,z$.

\textit{Nonequilibrium fluctuation-dissipation relations---} As shown below, Eqs.~(\ref{effpol1}) and (\ref{effpol2}) imply that FDT cannot hold simultaneously in the rest frame and in the lab frame. In equilibrium at zero temperature, FDT states that \cite{Landau5Book} $\eta^{(T=0)}_{jm}(\omega)= \,\mbox{sgn}(\omega)\,\mbox{Im}\left[ \alpha_{jm}(\omega)+\alpha_{mj}(\omega)\right]$. A direct application of FDT to the spinning sphere in the lab frame would yield a diagonal Hadamard tensor $\eta^{(\boldsymbol{\Omega})}(\omega)$, since $\alpha_{xy}^{(\boldsymbol{\Omega})}(\omega) = -\alpha_{yx}^{(\boldsymbol{\Omega})}(\omega)$ according to  Eq.~(\ref{effpol2}). However, using the rest frame isotropic Hadamard tensor $\xi(\omega) \equiv \eta(\omega)$ in Eq.~(\ref{effpol2}) leads to a nonzero off-diagonal Hadamard elements  $\eta_{xy}^{(\boldsymbol{\Omega})}(\omega)$.  This apparent contradiction arises because spinning drives the dipole degrees of freedom out of equilibrium in the lab frame. Invoking the FDT in the lab frame results in overlooking off-diagonal Hadamard terms that contribute significantly to the vdW interaction.




To find $\eta_{xy}^{(\boldsymbol{\Omega})}(\omega)$, we thus  assume that FDT holds in the rest frame only, and replace the resulting $\xi(\omega)$ into the right-hand side of Eq.~(\ref{effpol2}). We obtain the nonequilibrium result 
\begin{equation}
    \eta_{xy}^{(\boldsymbol{\Omega})}(\omega)=-2i\,\mbox{sgn}\,(\omega)\,\mbox{Re}\left[\alpha_{xy}(\omega)\right] \, , \label{FDTmod}
\end{equation}
when $|\Omega| < |\omega|$ and similar results otherwise~\cite{SM}. Equation~(\ref{FDTmod}) still links the fluctuations captured by $\eta^{(\boldsymbol{\Omega})}_{xy}$ with the dissipation in the system. However, the relation is not the same as the one given by FDT. The latter remains a good approximation in the quasi-static limit, where the angular frequencies are much smaller than the material's resonances. In this case, the modification of the vdW energy is weak, and the non-diagonal elements are negligible. 
However, we are particularly interested in the scenario where $\Omega$ is comparable to the resonance frequencies, highlighting the relevance of nonequilibrium physics for vdW interactions between spinning particles.

\textit{vdW interactions between spinning nanospheres---}Henceforth, we consider two nanospheres, denoted as $A$ and $B$, separated by a distance $R$. 
We assume that $R$ is much larger than the nanospheres' radii but much smaller than the characteristic wavelength associated with the material's relevant resonance (in the cm range for the example discussed later).
Within this range of distances, we may combine the dipole and non-retarded approximations.  
Our starting point is the corresponding expression for the vdW interaction energy between nanospheres $A$ and $B$~\cite{Santos2024}:
\begin{eqnarray}
    E & = & -\frac{\hbar\left(\delta_{jk}-3\hat{R}_j\hat{R}_k\right) \left(\delta_{mn}-3\hat{R}_m\hat{R}_n\right)} {128\pi^3\varepsilon^2_0 R^6} \label{londongeral} \nonumber \\
    & \times &  \int_{-\infty}^{\infty} \!\!\! d\omega \left[ \alpha^{A}_{jm}(\omega)\eta^{B*}_{kn}(\omega)+\eta^{A*}_{jm}(\omega)\alpha^{B}_{kn}(\omega) \right] , 
\end{eqnarray}
with an implicit sum on repeated indices. $\boldsymbol{R}$ denotes the position of object $B$ relative to object $A$, with $\boldsymbol{\hat{R}}=\boldsymbol{R}/R$. The nanospheres may rotate with angular velocities $\boldsymbol{\Omega}_A$ and $\boldsymbol{\Omega}_B$ along arbitrary directions, but we present below explicit results for two configurations of particular interest (other arrangements lead to similar results~\cite{SM}):  {\it (i)} both spheres rotate around the axis joining their centers ($\boldsymbol{\hat{R}}\parallel \boldsymbol{\hat{z}}$), represented as $\rightarrow\rightarrow$, and {\it (ii)} the spheres rotate parallel to each other but perpendicular to the line joining their centers ($\boldsymbol{\hat{R}}\parallel \boldsymbol{\hat{x}}$), represented as $\uparrow\uparrow$. The latter configuration could be implemented in a standard dual-beam optical trap in high vacuum~\cite{Ranjit2015, Rieser2022} (see also \cite{Jakubec2024} for a general theoretical description), employing circularly polarized trapping beams to drive the rotation~\cite{Monteiro2018, Reimann2018, Ahn2018, Ahn2020}. Regarding the $\rightarrow\rightarrow$ arrangement, a possible implementation would rely on chromatic aberration to align both nanospheres along the propagation direction \cite{Deplano2024}.

We obtain the interaction energy by substituting Eqs.~(\ref{effpol1}) and (\ref{effpol2}) into Eq.~(\ref{londongeral}), leading to
\begin{eqnarray}
    E_{\rightarrow \rightarrow} &=& 4[\mathcal{E}(\Omega_A-\Omega_B)+2\mathcal{E}(0)] \, , \label{rr}  \\ 
    E_{\uparrow \uparrow} &=& \mathcal{E}(\Omega_A-\Omega_B)+9\mathcal{E}(\Omega_A+\Omega_B)+2\mathcal{E}(0) \label{uu} \, .
\end{eqnarray}
Here, we define the auxiliary function
\begin{equation}
     \mathcal{E}(\Omega) = \mathcal{E}_{A\rightarrow B}(\Omega)+\mathcal{E}_{B\rightarrow A}(\Omega) \, . \label{ebaeab}
\end{equation}
The term
\begin{equation}
    \mathcal{E}_{B\rightarrow A}(\Omega) = -\frac{\mathcal{A}}{ R^6}\int_{-\infty}^{\infty}  d\omega [\alpha^{A}(\omega_+)+\alpha^{A}(\omega_-)]\eta^{B}(\omega)  \label{EBA},
\end{equation}
where $\mathcal{A} = \hbar/512\pi^3\varepsilon^2_0$, stands for the contribution due to the Doppler-shifted dipole induced in nanosphere $A$ due to dipole fluctuations in $B$, while
\begin{equation}
    \mathcal{E}_{A\rightarrow B}(\Omega) = -\frac{\mathcal{A}}{ R^6}\int_{-\infty}^{\infty}  d\omega [\eta^{A}(\omega_{+})+\eta^{A}(\omega_-)]\alpha^{B}(\omega) \label{EAB}
\end{equation}
describes the opposite: the correlation between Doppler-shifted dipole fluctuations in nanosphere $A$ and the induced response in $B$. Note that we can interchange the roles of $A$ and $B$ by performing a change of integration variable in Eqs.~(\ref{EBA}) and (\ref{EAB}). 

When $\boldsymbol{\Omega}_A=\boldsymbol{\Omega}_B=\boldsymbol{0}$, both expressions (\ref{rr}) and (\ref{uu}) reduce to $E^{(0)} = 12\mathcal{E}(0)$, which corresponds to the vdW interaction between nanospheres at rest~\cite{Santos2024}. Note that Eq.~(\ref{rr}) depends only on the relative angular velocity. This is expected for the $\rightarrow\rightarrow$ configuration, in which $A$ rotates with angular velocity $\Omega_A-\Omega_B$ from the perspective of $B$. The non-diagonal elements of $\eta_{ij}^\zeta$ ($\zeta = A, B$), absent if FDT were valid, play a pivotal role in this result. If equilibrium were assumed, $E_{\rightarrow\rightarrow}$ would not depend only on the relative angular velocity~\cite{SM}, highlighting the need for a full nonequilibrium treatment.

For the $\uparrow\uparrow$ configuration, the interaction energy depends not only on the relative velocity but also on $\Omega_A + \Omega_B$. This case lacks the symmetry of the previous one, explaining why the interaction energy is not solely a function of the relative velocity. The auxiliary function $\mathcal{E}(\Omega)$ is even, implying that the energy is invariant under the simultaneous exchange $\boldsymbol{\Omega}_A\rightarrow -\boldsymbol{\Omega}_A$ and $\boldsymbol{\Omega}_B\rightarrow -\boldsymbol{\Omega}_B$. This is consistent with energy being a scalar and angular velocity being a pseudovector quantity. It also implies that the results are symmetric under the exchange $A\leftrightarrow B$, as expected.

\textit{Lorentz model---}We can solve the integrals in Eqs.~(\ref{EBA}) and (\ref{EAB}) analytically when describing the polarizabilities by a Lorentz model. Before presenting a realistic numerical estimate, we 
first consider a simplified model with 
no dissipation, a single resonance, and zero temperature in order to gather physical insight. In this case, we have $\alpha^{\zeta}(\omega) = \alpha_{0\zeta} \omega_{0\zeta}^{2}/(\omega_{0\zeta}^{2}-\omega^2)$ and $\eta(\omega)=\pi\alpha_{0\zeta}\omega_{0\zeta} \left[\delta(\omega-\omega_{0\zeta})+\delta(\omega+\omega_{0\zeta})\right]$, where $\alpha_{0\zeta}$ and $\omega_{0\zeta}$ are the static polarizability and the internal resonance of the nanosphere $\zeta$, respectively. Substituting these expressions into Eq.~(\ref{EBA}) yields 
\begin{align}
    \mathcal{E}_{B\rightarrow A}&(\Omega) = -\frac{\hbar\alpha_{0A}\alpha_{0B}\omega_{0A}^2\omega_{0B}}{128\pi^2\varepsilon^2_0 R^6} \nonumber \\
    &\times \frac{\omega_{0A}^2-\omega_{0B}^2-\Omega^2}{[\Omega^2-(\omega_{0A}+\omega_{0B})^2][\Omega^2-(\omega_{0A}-\omega_{0B})^2]} \, . \label{ebaanalytic}
\end{align}
$\mathcal{E}_{A\rightarrow B}(\Omega)$ is obtained by exchanging $A\leftrightarrow B$ in the above equation. Normalizing this expression by its value without rotation, we obtain from Eqs.~(\ref{ebaeab}) and (\ref{ebaanalytic})
\begin{equation}
   \frac{ \mathcal{E}(\Omega) }{\mathcal{E}(0)}= \frac{(\omega_{0A}+\omega_{0B})^2}{(\omega_{0A}+\omega_{0B})^2-\Omega^2} \, .  \label{2levelsemdissip}
\end{equation}
Note that the ratio is independent of distance. Substituting this into Eq.~(\ref{rr}), we obtain the modification of the interaction energy for the $\rightarrow\rightarrow$ configuration. For simplicity, we assume identical nanospheres with $\omega_0\equiv \omega_{0A}=\omega_{0B}$ and $\alpha_0\equiv\alpha_{0A}=\alpha_{0B}$, which leads to
\begin{equation}
   \frac{ E_{\rightarrow\rightarrow}}{E^{(0)}} = \frac{1}{3}\left[\frac{4\omega_0^2}{4\omega_0^2-(\Omega_A-\Omega_B)^2}+2\right] \, . \label{rrlorentz}
\end{equation}
For $\Omega_A-\Omega_B=2\omega_0$, the interaction diverges, as the Lorentz model neglects dissipation. 
The $\uparrow\uparrow$ configuration is derived in an entirely analogous way and is given by
\begin{equation}
   \frac{ E_{\uparrow\uparrow}}{E^{(0)}} =\frac{\omega_0^2/3}{4\omega_0^2-(\Omega_A-\Omega_B)^2}+\frac{3\omega_0^2}{4\omega_0^2-(\Omega_A+\Omega_B)^2}+\frac{1}{6} \, . \label{uulorentz}
\end{equation}
In the high spinning frequency limit $\Omega_A \gg 2\omega_0$ (with nanosphere $B$ at rest), the $\{\uparrow \uparrow,\rightarrow \rightarrow\}$ arrangements exhibit distinct behaviors. In this regime, the response of nanosphere $A$ transverse to the rotation axis probes the material transparency limit ($\alpha(\omega)\rightarrow 0$ when $\omega \rightarrow\infty$), so that only the longitudinal contribution from $\alpha_{zz}^{(A)}(\omega)$ survives. The same holds for the Hadamard Green's function. The fast-spinning nanosphere is thus akin to a quantum needle polarizable only along its rotation axis. In this limit, vdW interaction energies (and forces) are 4 times larger in the  $\uparrow \uparrow$ ($E_{\uparrow \uparrow} \rightarrow 2E^{(0)}/3$) than in the $\rightarrow \rightarrow$ ($E_{\rightarrow \rightarrow}  \rightarrow E^{(0)}/6$) arrangement, respectively orthogonal and parallel to the needle axis. A stronger suppression of vdW forces occurs in the latter case. 

In the following, we present results for a realistic material at a 
finite temperature $T$. We confirm the main features discussed in connection with the simpler analytical model presented above, which suggests that rotation can indeed have a marked effect on the vdW interaction.

\textit{vdW interaction between spinning BST nanospheres---}As a concrete application, we consider nanospheres composed of barium strontium titanate (BST). This material is advantageous due to its low polaritonic frequency in the GHz range~\cite{Xu2021}. The electric permittivity of BST can be described by the Lorentz-Drude model as a sum of harmonic oscillators corresponding to various internal resonances. For our purposes, the most critical aspect is accurately modeling the peak associated with the polaritonic resonance. Then, we adopt the permittivity
\begin{equation}
    \frac{\varepsilon(\omega)}{\varepsilon_0}=1+\frac{f_0\tilde{\omega}_0^2}{\tilde{\omega}_0^2-\omega^2-i\gamma_0\omega}  \, , \label{permittivity}
\end{equation}
where $f_0=12.2$, $\tilde{\omega}_0=5.7 \times 10^9$ Hz, and $\gamma_0= 2.8 \times 10^8\,{\rm Hz}$ accounts for dissipation \cite{Xu2021, Turky2015}. Using Eq.~(\ref{permittivity}), the polarizability of a BST nanosphere of radius $a$ is given by \cite{JacksonBook}
\begin{equation}
    \alpha(\omega) = \frac{4\pi\varepsilon_0 a^3f_0\tilde{\omega}_0^2}{3(\omega_0^2-\omega^2-i\gamma_0\omega)} \, ,
\end{equation}
where $\omega_0=\tilde{\omega}_0\sqrt{1+f_0/3}$ represents the polaritonic resonance frequency of the nanosphere in the dipole approximation. For nanospheres at rest, the Hadamard functions are obtained through the FDT relation and are given by $\eta^{(T)}(\omega)=2\coth(\hbar\omega/2k_BT)\,\mbox{Im}\,\alpha(\omega)$.

\begin{figure}[b]
\includegraphics[scale=0.45]{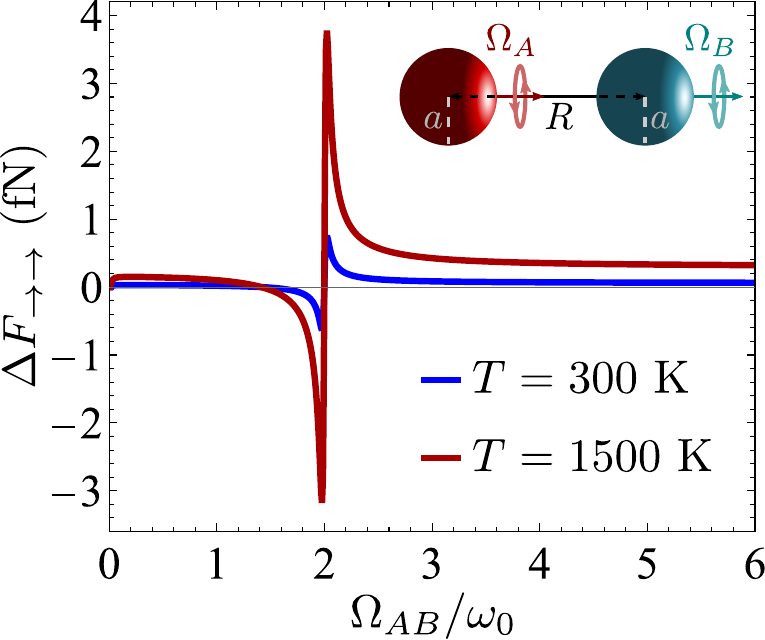}
\caption{Modification of the vdW force for the $\rightarrow\rightarrow$ configuration as a function of the relative angular velocity $\Omega_{AB}=\Omega_A-\Omega_B$ (in units of the polaritonic resonance frequency $\omega_0$). Positive values indicate a repulsive contribution. The nanospheres are made of BST, with material parameters given in the main text. }
\label{Fig:rr}
\end{figure}

\begin{figure}[t]
\includegraphics[scale=0.32]{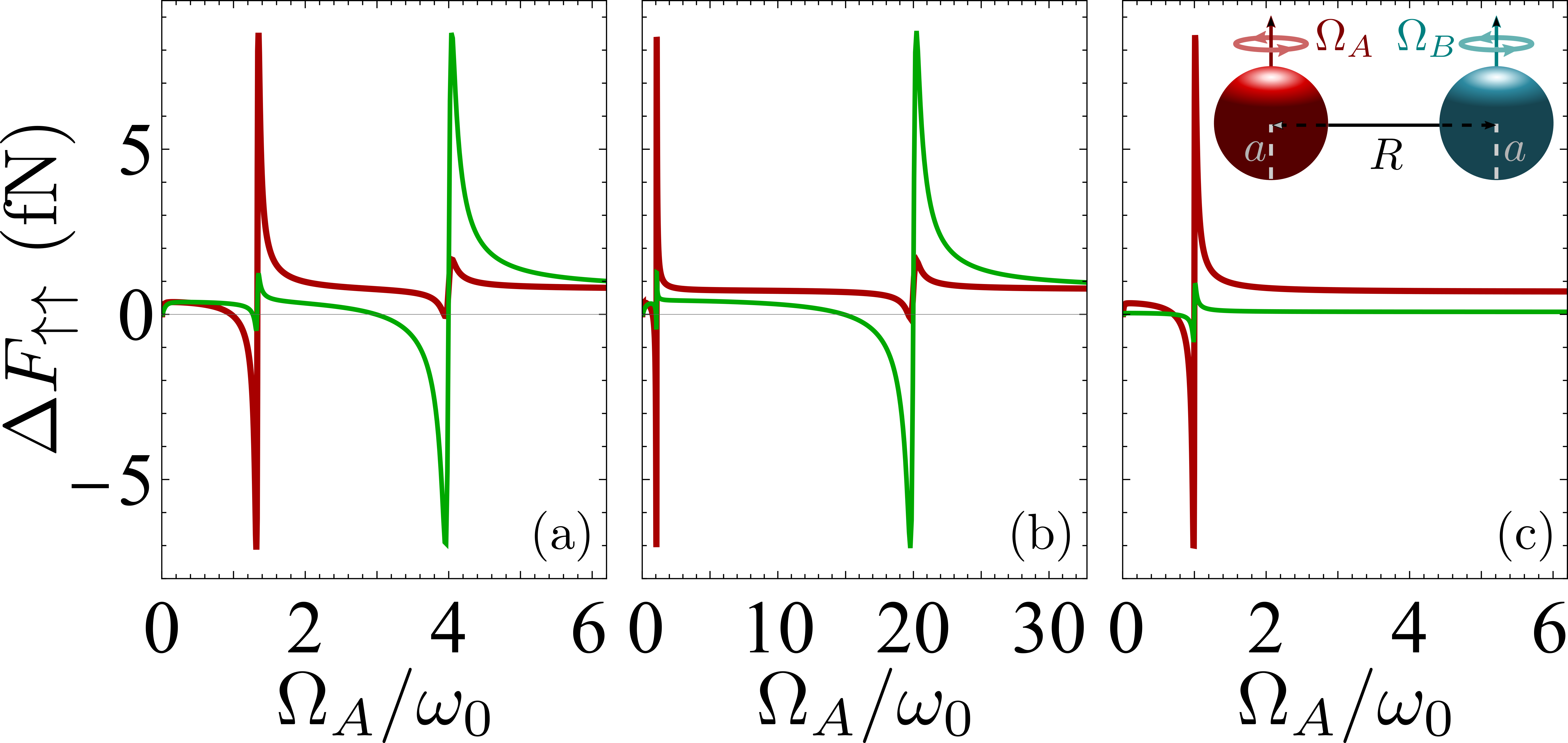}
\caption{Modification of the vdW force for co-rotating (red) and counter-rotating (green) BST nanospheres in the $\uparrow\uparrow$ configuration as a function of $\Omega_A$ (in units of the polaritonic resonance frequency $\omega_0$) for $|\Omega_B|/|\Omega_A| $ equal to (a) $0.5$, (b) $0.9$, and (c) $1$. We take $T=1500\,{\rm K.}$}
\label{Fig:uu}
\end{figure}

We now evaluate the rotation-induced vdW force between spinning nanospheres for the $\{\uparrow \uparrow,\rightarrow \rightarrow\}$ configurations. We numerically evaluate the integrals given in Eqs.~(\ref{EBA}) and (\ref{EAB}) and then substitute the results into Eqs.~(\ref{rr}) and (\ref{uu}) to calculate the modification of the force arising from rotation. We plot the results $\Delta{F}_{\rightarrow\rightarrow}$ and $\Delta{F}_{\uparrow\uparrow}$ as functions
of the relevant rotation frequency in Figs.~\ref{Fig:rr} and \ref{Fig:uu}, respectively.
We take $a=60$~nm and $R=180$~nm for the radius and distance between the nanospheres.

Although the magnitude of the force modification strongly depends on temperature, the main qualitative features appearing in Figs.~~\ref{Fig:rr} and \ref{Fig:uu} can be understood in terms of the simplified zero-temperature model discussed earlier. 
In Fig.~\ref{Fig:rr}, we plot results for the $\rightarrow\rightarrow$ configuration at $T=300\,{\rm K}$ (blue) and $T=1500\,{\rm K}$ (red). From Eq.~(\ref{rrlorentz}), we expect a single resonance peak at $\Omega_A=2\omega_0+\Omega_B$  for the $\rightarrow\rightarrow$ arrangement, which agrees with Fig.~\ref{Fig:rr}. Also in qualitative agreement with Eq.~(\ref{rrlorentz}), a strong repulsive contribution (positive values of $\Delta F_{\rightarrow\rightarrow}$) 
emerges on the high-frequency side of the resonance peak, and $\Delta F_{\rightarrow\rightarrow}$ saturates at high rotation frequencies. Temperature 
mainly modifies the saturation as well as the peak values. 
Notably, increasing temperature also leads to a small repulsive contribution at low rotation frequencies.

Even stronger effects are observed in the $\uparrow\uparrow$ configuration, which could be implemented with standard dual-beam optical traps~\cite{Ranjit2015, Rieser2022}. The force modification in this configuration is shown in Fig.~\ref{Fig:uu} for co-rotating (red curve) and counter-rotating (green curve) nanospheres. 
As the internal temperature of an optically levitated nanosphere in vacuum is typically high due to laser absorption~\cite{Gonzalez-Ballestero2021}, we take $T=1500\,{\rm K}$.
The magnitude of the repulsive contribution reaches $8.5\,{\rm fN}.$ At room temperature (not shown), peak values are at the $\rm fN$ level. 
%
The total vdW force is highly sensitive to the material's electromagnetic response in the UV range, which has not been characterized for BST. Nonetheless, taking typical values for the Hamaker constant of several dielectric materials~\cite{IsraelachviliBook} (around $5\times 10^{-20}$ J), we estimate the (attractive) vdW force between non-spinning nanospheres to be $4\,{\rm fN}$~\cite{SM} for the geometric parameters considered in Fig.~\ref{Fig:uu}. This indicates that the strong repulsive contribution from rotation, which concerns only the polaritonic resonance, could overcome the attractive contribution from UV resonances (which is not affected by a GHz rotation), thus resulting in an overall repulsive vdW force.

The comparison between Figs.~\ref{Fig:rr} and \ref{Fig:uu} clearly shows that the $\uparrow\uparrow$ configuration might contain an additional resonance peak. Indeed, the interaction energy in this case also involves the auxiliary function $\mathcal{E}$ evaluated at $\Omega_{A}+\Omega_B$, leading to a peak at $\Omega_A=2\omega_0-\Omega_B$ which is missing in the $\rightarrow\rightarrow$ case by symmetry. 
As $|\Omega_B|$ approaches $|\Omega_A|$, the additional peak shifts to higher frequencies, as depicted in Fig.~\ref{Fig:uu}(b), leaving a single peak at $|\Omega_A|=\omega_0$ when $|\Omega_B|=|\Omega_A|,$  as can be seen in Fig.~\ref{Fig:uu}(c). The co-rotating and counter-rotating cases are related by an exchange between the peak intensities~[Figs.~\ref{Fig:uu}(a) and \ref{Fig:uu}(b)], as follows from Eq.~(\ref{uu}). For the co-rotating configuration, the strongest peak occurs at a lower frequency.

The striking results presented here arise directly from the rotational Doppler shift. To elucidate this connection, we first consider the case without rotation ($\Omega=0$). Even in this scenario, the contribution $\mathcal{E}_{B\rightarrow A}$ given in Eq.~(\ref{ebaanalytic}) diverges when $\omega_{0A}=\omega_{0B}$, which is expected due to resonance when dissipation is neglected. However, only the sum $\mathcal{E}=\mathcal{E}_{A\rightarrow B}+\mathcal{E}_{B\rightarrow A}$ is physically meaningful and remains finite when $\Omega=0$, including at $\omega_{0A}=\omega_{0B}$, as can be checked in Eq.~(\ref{2levelsemdissip}). 
This cancellation of divergences occurs because, when $\Omega=0$, the contributions $\mathcal{E}_{B\rightarrow A}$ and $\mathcal{E}_{B\rightarrow A}$ have opposite signs for any $\omega_{0A}\neq \omega_{0B}$. 
This means that if one is attractive, the other is repulsive. The latter arises because the slower material (with the smaller transition frequency) cannot follow the dipole fluctuations of the faster one, implying that the induced dipole is opposite to the electric field acting on it (the polarizability becomes negative at frequencies higher than the resonance of the material)~\cite{Santos2024}. In the static case, the repulsive contribution is weaker than the attractive one, yielding the usual attractive vdW force. 

With rotation, we can engineer each contribution, disrupting the near-perfect cancellation coming from resonant contributions of opposite signs. For simplicity, consider identical nanospheres, where Eq.~(\ref{ebaanalytic}) simplifies to $\mathcal{E}_{B\rightarrow A}=-\hbar\alpha_0^2\omega_0^3/[128\pi^2\varepsilon_0^2R^6(4\omega_0^2-\Omega^2)]=\mathcal{E}_{A\rightarrow B}$. Both contributions now have the same sign. They are attractive for $\Omega<2\omega_0$ and repulsive otherwise, exactly as observed in our results. This repulsion can be intuitively understood as follows: consider nanosphere $A$ rotating with angular velocity $\Omega$ while $B$ is at rest. 
Due to the Doppler shift, the dipole of $A$ fluctuates with frequencies $\omega_{\pm}=\omega_0\pm\Omega$. When $\Omega > 2\omega_0$, we have $|\omega_{\pm}|>\omega_0$, showing that nanosphere $B$ cannot follow the fast oscillations of $A$. In addition, the oscillations of $B$ with frequency $\omega_0$ are perceived by $A$ as occurring at frequencies $|\omega_{\pm}|>\omega_0$, explaining that $A$ cannot follow $B$ either. 
Hence, when $\Omega > 2\omega_0$, both nanospheres fail to follow the fluctuations of the other, resulting in repulsion. The fact that both contributions, $A\rightarrow B$ and $B\rightarrow A$, have the same sign is the key element to the stronger results we obtain. Now, the resonances add instead of subtracting. The case $\omega_{0A}\neq\omega_{0B}$ can be understood analogously.

In conclusion, we have demonstrated that rotation can profoundly impact the vdW interaction between nanospheres. It can enhance attraction, suppress it, and possibly reverse it into repulsion. Rotation is therefore a promising mechanism for controlling vdW forces. As the rotation frequency is proportional to the trapping beam power (and also depends critically on the pressure of the vacuum chamber), it can be readily used as a knob for different applications. For example, the dependence on $\Omega_A$ and $\Omega_B$ can be leveraged to isolate the vdW interaction from the electrostatic and optical binding ones, allowing for better control over the interactions between particles in levitodynamics. 

Our results can be extended to diverse geometries. Perspectives for this work also include tailoring vdW interactions in systems where rotational degrees of freedom are quantized, such as in molecular physics or quantum optomechanics~\cite{Bang2020, Gao2024, Stickler2021}.

\begin{acknowledgments}

We thank Guilherme Matos, Wilton Kort-Kamp, Pedro Pereira, Thomas Zanon, Daniel Jonathan, Carlos Farina, and Lucas Sigaud for fruitful discussions. F.I., P.A.M.N., and R.M.S. were partially supported by Conselho Nacional de Desenvolvimento Cient\'{\i}fico e Tecnol\'ogico (CNPq), Coordenaç\~ao de Aperfeiçamento de Pessoal de N\'{\i}vel Superior (CAPES),  Instituto Nacional de Ci\^encia e Tecnologia de Fluidos Complexos  (INCT-FCx), and the Research Foundations of the States of Rio de Janeiro (FAPERJ) and S\~ao Paulo (FAPESP). H.S.G.A. thanks FAPERJ and CAPES for funding. P.P.A. acknowledges funding from CNPq (Grant No. 152050/2024-8). 

\end{acknowledgments}

\bibliography{SpinningCasimir}


\newpage


\begin{widetext}

\section*{Supplemental Material: Tailoring the van der Waals interaction with rotation}


This Supplemental Material provides additional details on (I) the derivation of the dipole correlation functions of the spinning nanospheres, (II) the evaluation of the van der Waals (vdW) interaction energy for several geometric arrangements of the two rotation axes and relative positions of the spinning nanospheres, (III) the total vdW force between nanospheres at rest, and (IV) the inconsistencies arising from the assumption of thermal equilibrium in the rotating frame.

\section{Dipole response functions in a rotating frame}

Let us consider a nanosphere spinning around the $z-$axis with constant angular velocity $\boldsymbol{\Omega}=\Omega \boldsymbol{\hat{z}}$. We assume the Hamiltonian in the rotating frame is the rest Hamiltonian $H^{(0)}$. We denote the Hamiltonian in the inertial laboratory frame by $H^{(\boldsymbol{\Omega})}$, where the index $\boldsymbol{\Omega}$ indicates that the Hamiltonian in the laboratory frame is dependent on the angular velocity of the nanosphere. The relation between the rest (rotating) frame Hamiltonian $H^{(0)}$ and the laboratory frame is given by\cite{Dalibard05,RMPFetter} 
\begin{equation}
    H^{(\boldsymbol{0})}=H^{(\boldsymbol{\Omega})}- \boldsymbol{\Omega} \cdot \boldsymbol{L}\, .
\end{equation}
The dipole operator in the Heisenberg picture in the laboratory frame is given by
\begin{equation}
    \boldsymbol{d}^{(\boldsymbol{\Omega})}(t) = e^{\frac{i}{\hbar}H^{(\boldsymbol{\Omega})} t}\boldsymbol{d}e^{-\frac{i}{\hbar}H^{(\boldsymbol{\Omega})}t} = e^{\frac{i}{\hbar}(H^{(0)} +\boldsymbol{\Omega} \cdot \boldsymbol{L})t}\boldsymbol{d}e^{-\frac{i}{\hbar}(H^{(0)}+\boldsymbol{\Omega} \cdot \boldsymbol{L})t}\, .
\end{equation}
%

%
Spherical symmetry guarantees that $[H^{(0)},L_z]=0$, which allows us to write
\begin{equation}
    \boldsymbol{d}^{(\boldsymbol{\Omega})}(t) = e^{\frac{i}{\hbar}L_z\Omega t}\boldsymbol{d}^{(0)}(t)e^{-\frac{i}{\hbar}L_z\Omega t} \,,
\end{equation}
where $\boldsymbol{d}^{(0)}(t)$ denotes the dipole operator in the Heisenberg picture for the non-rotating case. Explicitly evaluating this expression and decomposing it in components with respect to the lab inertial frame, we obtain
\begin{eqnarray}
    d_{x}^{(\boldsymbol{\Omega})}(t)&=&  d_{x}^{(0)}(t)\cos(\Omega t)-d_{y}^{(0)}(t)\sin (\Omega t) \, ,  \\
    d_{y}^{(\boldsymbol{\Omega})}(t)&=& d_{y}^{(0)}(t)\cos(\Omega t)+d_{x}^{(0)}(t)\sin (\Omega t) \, , \\
    d_{z}^{(\boldsymbol{\Omega})}(t)&=&d_{z}^{(0)}(t)\, . \label{componentesdRprime}
\end{eqnarray}

Assuming the electronic state co-rotates with the nanosphere -- which is a good approximation here since we assume $\Omega$ to be much smaller than the electronic characteristic frequency (typically at $10^{15}-10^{16}$ Hz) --, the correlations $\langle d^{(\boldsymbol{\Omega})}_i(t)d^{(\boldsymbol{\Omega})}_j(t')\rangle$ can be readily related to their counterparts in the absence of rotation. Due to the spherical symmetry of $H^{(0)}$, $\langle d_{i}^{(0)}(t)d^{(0)}_{j}(t')\rangle=\langle \boldsymbol{d}^{(0)}(t)\cdot\boldsymbol{d}^{(0)}(t')\rangle \delta_{ij}/3$, yielding
\begin{eqnarray}
    \langle d_{x}^{(\boldsymbol{\Omega})}(t)d_{x}^{(\boldsymbol{\Omega})}(t')\rangle &=&\frac{\langle \boldsymbol{d}^{(0)}(t)\cdot\boldsymbol{d}^{(0)}(t')\rangle}{3}\cos[\Omega(t-t')]= \langle d_{y}^{(\boldsymbol{\Omega})}(t)d_{y}^{(\boldsymbol{\Omega})}(t')\rangle \,, \\
    \langle d_{x}^{(\boldsymbol{\Omega})}(t)d_{y}^{(\boldsymbol{\Omega})}(t')\rangle &=&-\frac{\langle \boldsymbol{d}^{(0)}(t)\cdot\boldsymbol{d}^{(0)}(t')\rangle}{3}\sin[\Omega(t-t')]= -\langle d_{y}^{(\boldsymbol{\Omega})}(t)d_{x}^{(\boldsymbol{\Omega})}(t')\rangle \,, \\
    \langle d_{i}^{(\boldsymbol{\Omega})}(t)d_{z}^{(\boldsymbol{\Omega})}(t')\rangle &=&\frac{\langle \boldsymbol{d}^{(0)}(t)\cdot\boldsymbol{d}^{(0)}(t')\rangle}{3} \delta_{iz}= \langle d_{z}^{(\boldsymbol{\Omega})}(t)d_{i}^{(\boldsymbol{\Omega})}(t')\rangle \, .
\end{eqnarray}
These expressions determine the polarizability tensor in the rotating frame, defined as $\alpha^{(\boldsymbol{\Omega})}_{ij}=(i/\hbar)\theta(t-t')\langle[d_i(t),d(t')]\rangle$. When $\Omega=0$, spherical symmetry ensures $\alpha_{ij}^{(0)}=\alpha\delta_{ij}$, leading to
\begin{eqnarray}
    \alpha_{xx}^{(\boldsymbol{\Omega})}(t-t')&=&\alpha_{yy}^{(\boldsymbol{\Omega})}(t-t')=\alpha(t-t')\cos[\Omega(t-t')]\,, \\ 
    \alpha_{xy}^{(\boldsymbol{\Omega})}(t-t')&=&-\alpha_{yx}^{(\boldsymbol{\Omega})}(t-t')=-\alpha(t-t')\sin[\Omega(t-t')] \,,\\ 
    \alpha_{jz}^{(\boldsymbol{\Omega})}(t-t')&=&\alpha_{zj}^{(\boldsymbol{\Omega})}(t-t')=\alpha(t-t')\delta_{jz}
    \, .
\end{eqnarray}
These components satisfy $\alpha_{ij}^{(\boldsymbol{\Omega})}=-\alpha_{ji}^{(\boldsymbol{\Omega})}$, as required by general symmetry considerations \cite{Landau5Book}. At equal times $t=t'$, the polarizability tensor $\alpha_{ij}^{(\boldsymbol{\Omega})}$ transforms like a rank-2 tensor under rotations and remains proportional to the identity, as expected. However, for $t\neq t'$, this no longer holds, as each vector $\boldsymbol{d}$ at different times undergoes distinct rotation.

We also need the transformation rule for the symmetric correlation function, defined by $\eta_{ij}^{(\boldsymbol{\Omega})}=(1/\hbar)\{d_i(t),d_j(t')\}$. Similar considerations yield 
\begin{eqnarray}
    \eta_{xx}^{(\boldsymbol{\Omega})}(t-t')&=&\eta_{yy}^{(\boldsymbol{\Omega})}(t-t')=\eta(t-t')\cos[\Omega(t-t')] \,, \\
    \eta_{xy}^{(\boldsymbol{\Omega})}(t-t')&=&-\eta_{yx}^{(\boldsymbol{\Omega})}(t-t')=-\eta(t-t')\sin[\Omega(t-t')] \,, \\ 
      \eta_{jz}^{(\boldsymbol{\Omega})}(t-t')&=&\eta_{zj}^{(\boldsymbol{\Omega})}(t-t')=\eta(t-t')\delta_{jz}
    \, . \label{etaOmega}
\end{eqnarray}
In Fourier space, the effect of rotation manifests as frequency shifts in the Fourier components
\begin{eqnarray}
    \eta_{xx}^{(\boldsymbol{\Omega})}(\omega)&=&\frac{\eta(\omega_+)+\eta(\omega_-)}{2} =\eta_{yy}^{(\boldsymbol{\Omega})}(\omega)  \label{etaFourierOmegaxx} \,,\\ 
    \eta_{xy}^{(\boldsymbol{\Omega})}(\omega)&=&\frac{i[\eta(\omega_+)-\eta(\omega_-)]}{2} =-\eta_{yx}^{(\boldsymbol{\Omega})}(\omega)  \label{etaFourierOmegaxy} \,,\\
    \eta_{jz}^{(\boldsymbol{\Omega})}(\omega)&=&\eta(\omega)\delta_{jz} = \eta_{zj}^{(\boldsymbol{\Omega})}(\omega)\, , \label{etaFourierOmega}
\end{eqnarray}
with $\omega_{\pm}=\omega\pm\Omega$. Because the diagonal components of $\eta^{(\boldsymbol{\Omega})}_{ij}(t-t')$  are real and even functions of $t-t'$, their Fourier transforms are also real and even in $\omega$. Our results satisfy these expected constraints, as can be demonstrated using that, in the absence of rotation, $\eta(\omega)$ satisfies them.  In contrast, the off-diagonal component $\eta_{xy}^{(\boldsymbol{\Omega})}(t-t')$ is real odd function, implying $\eta_{xy}^{(\boldsymbol{\Omega})}(\omega)$ to be purely imaginary and odd in $\omega$. The same symmetry considerations apply to the polarizability
\begin{eqnarray}
    \alpha_{xx}^{(\boldsymbol{\Omega})}(\omega)&=&\frac{\alpha(\omega_+)+\alpha(\omega_-)}{2} =\alpha_{yy}^{(\boldsymbol{\Omega})}(\omega) \label{alphaFourierOmega1} \,, \\ 
    \alpha_{xy}^{(\boldsymbol{\Omega})}(\omega)&=&\frac{i[\alpha(\omega_+)-\alpha(\omega_-)]}{2} =-\alpha_{yx}^{(\boldsymbol{\Omega})}(\omega) \label{alphaFourierOmega2} \,, \\
     \alpha_{jz}^{(\boldsymbol{\Omega})}(\omega)&=&\alpha(\omega)\delta_{jz} = \alpha_{zj}^{(\boldsymbol{\Omega})}(\omega) \, .
    \label{alphaFourierOmega3}
\end{eqnarray}
The polarizability is constrained such that its real part is even while its imaginary part is odd in $\omega$, which follows from the reality of $\alpha^{(\boldsymbol{\Omega})}_{ij}(t-t')$. Assuming this property holds for $\alpha(\omega)$, it follows from Eqs.~(\ref{alphaFourierOmega1})-(\ref{alphaFourierOmega3}) that it remains valid for the spinning nanosphere, as it should. 

It may be directly verified that Eqs.~(\ref{etaFourierOmegaxx})-(\ref{alphaFourierOmega3}) obey the fluctuation-dissipation theorem (FDT), which states that, at zero temperature, 
\begin{equation}
\eta^{(T=0)}_{jm}(\omega)= \,\mbox{sgn}(\omega)\,\mbox{Im}\left[ \alpha_{jm}(\omega)+\alpha_{mj}(\omega)\right] \,, \label{FDT} \end{equation}
even if in the absence of rotation $\alpha(\omega)$ and $\eta(\omega)$ satisfy such a relation. Assuming the validity of FDT at zero temperature for the static case, we can substitute $\eta(\omega)=2\mbox{sgn}(\omega)\alpha(\omega)$ in the right-hand side of Eq.~(\ref{etaFourierOmegaxx}), obtaining the modified and nonequilibrium FDT relations
 \begin{equation}
  \eta_{xx}^{(\boldsymbol{\Omega})}(\omega) =    \left\{\begin{matrix}
2\,\mbox{sgn}(\omega)\,\mbox{Im}[\alpha_{xx}^{(\boldsymbol{\Omega})}(\omega)], \;\; \mbox{if}\;\; |\omega|\geq|\Omega| \, ,\\
 -2\,\mbox{sgn}(\Omega)\,\mbox{Re}[\alpha^{(\boldsymbol{\Omega})}_{xy}(\omega)], \;\; \mbox{if}\;\; |\omega|\leq|\Omega| \, ,
\end{matrix}\right. 
 \end{equation}
with analogous relation for $\eta_{yy}^{(\boldsymbol{\Omega})}(\omega)$. It should be noted that the equilibrium FDT relation is valid only for $|\omega|>\Omega$. Recalling that $\alpha(0)$ is real, we can check that $\eta_{xx}$ is a continuous function. Finally, for the off-diagonal component:
 \begin{equation}
  \eta_{xy}^{(\boldsymbol{\Omega})}(\omega) =    \left\{\begin{matrix}
-2i\,\mbox{sgn}(\omega)\,\mbox{Re}[\alpha_{xy}^{(\boldsymbol{\Omega})}(\omega)], \;\; \mbox{if}\;\; |\omega|\geq|\Omega| \, ,\\
 2i\,\mbox{sgn}(\Omega)\,\mbox{Im}[\alpha_{xx}^{(\boldsymbol{\Omega})}(\omega)], \;\; \mbox{if}\;\; |\omega|\leq|\Omega| \, .
\end{matrix}\right.
 \end{equation}
 Here, the standard equilibrium FDT relations are recovered only in the absence of rotation ($\Omega = 0$). When the non-spinning sphere is a at temperature $T$, we may also derive modified FDT relations by employing $\eta(\omega)=2\coth(\hbar\omega/2k_BT) \, \alpha(\omega)$. Defining $f(\omega,T)=(1/2)[\coth(\hbar(\omega-\Omega)/2k_BT)+\coth(\hbar(\omega+\Omega)/2k_BT)]$ and $g(\omega,T)=(1/2)[\coth(\hbar(\omega-\Omega)/2k_BT)-\coth(\hbar(\omega+\Omega)/2k_BT)]$, we obtain
 \begin{eqnarray}
       \eta_{xx}^{(\boldsymbol{\Omega})}(\omega) &=& 2f(\omega,T)\,\mbox{Im}\left[\alpha^{(\boldsymbol{\Omega})}_{xx}(\omega) \right]+2g(\omega,T)\,\mbox{Re}\left[\alpha_{xy}^{(\boldsymbol{\Omega})}(\omega) \right] \, , \\
        \eta_{xy}^{(\boldsymbol{\Omega})}(\omega) &=& -2if(\omega,T)\,\mbox{Re}\left[\alpha^{(\boldsymbol{\Omega})}_{xy}(\omega)\right]-2ig(\omega,T)\,\mbox{Im}\left[\alpha^{(\boldsymbol{\Omega})}_{xx}(\omega)\right] \, . 
 \end{eqnarray}
 When $T=0$, $f=\,\mbox{sgn}(\omega)$ and $g=0$, for $|\Omega|<|\omega|$, and $f=0$ and $g=-\,\mbox{sgn}(\Omega)$, for $|\Omega|>|\omega|$, recovering the previous results.

\section{Interaction energy between rotating nanospheres for different geometric arrangements}

We investigate below the interaction energy for several different geometric configurations of spinning nanospheres. Each geometric arrangement is determined by fixing both the direction of each nanosphere’s rotation axis and the line joining their center-of-mass positions.

The main effect of rotation is to replace the spheres' rest response functions with effective anisotropic ones, given by Eqs.~(\ref{etaFourierOmegaxx})-(\ref{alphaFourierOmega3}). In the dipole approximation, the van der Waals interaction between anisotropic bodies is described by~\cite{Santos2024}
\begin{equation}
    E = -\frac{\hbar\left(\delta_{jk}-3\hat{R}_j\hat{R}_k\right)\left(\delta_{mn}-3\hat{R}_m\hat{R}_n\right)}{128\pi^3\varepsilon^2_0 R^6}\int_{-\infty}^{\infty} d\omega \left[ \alpha^{A(\boldsymbol{\Omega}_A)}_{jm}(\omega)\eta^{*B(\boldsymbol{\Omega}_B)}_{kn}(\omega)+\eta^{*A(\boldsymbol{\Omega}_A)}_{jm}(\omega)\alpha^{A(\boldsymbol{\Omega}_B)}_{kn}(\omega) \right] . \label{londongeral}
\end{equation}
To explicitly evaluate this expression, we now examine some particular configurations of the rotating bodies, as depicted in Fig.~\ref{Fig:configurations}.

\begin{figure}[h]
\includegraphics[scale=0.9]{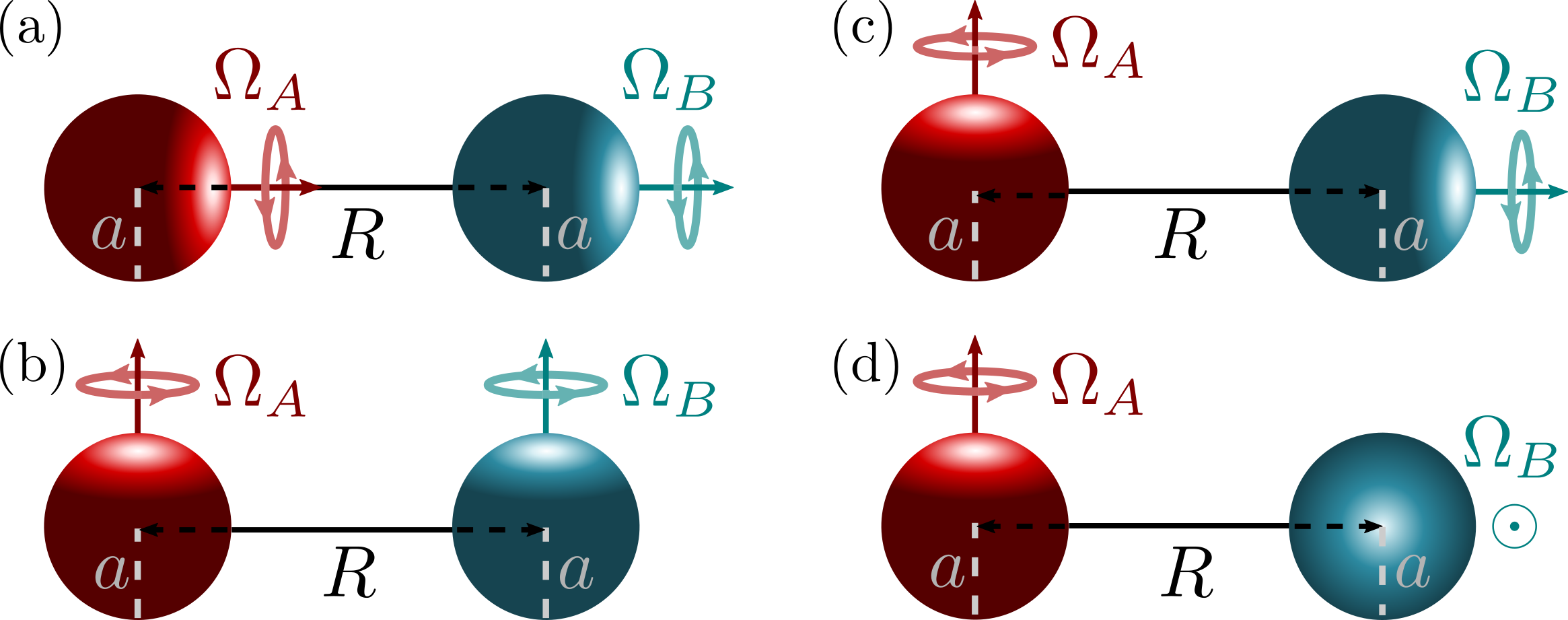}
\caption{Different geometric arrangements of the two rotation axes and relative positions of the spinning nanospheres.}
\label{Fig:configurations}
\end{figure}

\subsection{Spheres spinning parallel to each other and parallel to the line joining them $(\rightarrow\rightarrow)$}

In this configuration, shown in Fig.~\ref{Fig:configurations}(a), we can assume $\boldsymbol{\Omega}_\zeta=\Omega_\zeta\boldsymbol{\hat{z}}$ ($\zeta=A,B$) and $\boldsymbol{R}=R\boldsymbol{\hat{z}}$. Due to this symmetry, the $xz,yz,zx,zy$ components of the response functions vanish, allowing us to simplify Eq.~(\ref{londongeral}). Labeling this configuration with the subscript $\rightarrow\rightarrow$, we obtain
\begin{equation}
    E_{\rightarrow\rightarrow} = -\frac{\hbar}{64\pi^3\varepsilon^2_0 R^6}\int_{-\infty}^{\infty} \!\!\! d\omega \left[ \alpha^{A(\boldsymbol{\Omega}_A)}_{xx}(\omega)\eta^{B(\boldsymbol{\Omega}_B)}_{xx}(\omega)-  \alpha^{A(\boldsymbol{\Omega}_A)}_{xy}(\omega)\eta^{B(\boldsymbol{\Omega}_B)}_{xy}(\omega)+ 2\alpha^{A(\boldsymbol{\Omega}_A)}_{zz}(\omega)\eta^{B(\boldsymbol{\Omega}_B)}_{zz}(\omega)+A\leftrightarrow B\right] , \label{errnonequilibriumgeneral}
\end{equation}
where we used that $\alpha^{A(\boldsymbol{\Omega}_A)}_{yy}\eta^{B(\boldsymbol{\Omega}_B)}_{yy}=\alpha_{xx}^{A(\boldsymbol{\Omega}_A)}\eta^{B(\boldsymbol{\Omega}_B)}_{xx}$ and $\alpha^{A(\boldsymbol{\Omega}_A)}_{xy}\eta^{B(\boldsymbol{\Omega}_B)}_{xy}=\alpha_{yx}^{A(\boldsymbol{\Omega}_A)}\eta^{B(\boldsymbol{\Omega}_B)}_{yx}$. We also used that $\eta_{xx}$ is a real number while $\eta_{xy}$ is purely imaginary. Substituting the expressions from Eqs.~(\ref{etaFourierOmegaxx})-(\ref{alphaFourierOmega3}), the interaction energy becomes
\begin{equation}
    E_{\rightarrow\rightarrow} = -\frac{\hbar}{128\pi^3\varepsilon^2_0 R^6}\int_{-\infty}^{\infty} \!\!\! d\omega \left[ \alpha^{A}(\omega_{A+})\eta^{B}(\omega_{B+})+  \alpha^{A}(\omega_{A-})\eta^{B}(\omega_{B-})+ 4\alpha^{A}(\omega)\eta^{B}(\omega)+A\leftrightarrow B\right] \label{ERRalphaeta} ,
\end{equation}
with $\omega_{\zeta \pm}=\omega\pm\Omega_\zeta$. By changing variables so that the information on the Doppler shift is entirely contained in the response functions of nanosphere $A$, we may recast the above integrals in the form
\begin{equation}
     E_{\rightarrow \rightarrow} = 4[\mathcal{E}(\Omega_A-\Omega_B)+2\mathcal{E}(0)] \,, \label{smrr}
\end{equation}
where $\mathcal{E}(\Omega)$ is the auxiliary function introduced in the main text.

\subsection{Spheres spinning parallel to each other and perpendicular to the line joining them $(\uparrow\uparrow)$}

In this case of Fig.~\ref{Fig:configurations}(b), we can assume $\boldsymbol{\Omega}_\zeta=\Omega_\zeta\boldsymbol{\hat{z}}$ and $\boldsymbol{R}=R\boldsymbol{\hat{x}}$. Similar calculations to those performed in the previous section simplify Eq.~(\ref{londongeral}) to
\begin{equation}
    E_{\uparrow\uparrow} = -\frac{\hbar}{128\pi^3\varepsilon^2_0 R^6}\int_{-\infty}^{\infty} \!\!\! d\omega \left[ 5\alpha^{A(\boldsymbol{\Omega}_A)}_{xx}(\omega)\eta^{B(\boldsymbol{\Omega}_B)}_{xx}(\omega)+  4\alpha^{A(\boldsymbol{\Omega}_A)}_{xy}(\omega)\eta^{B(\boldsymbol{\Omega}_B)}_{xy}(\omega)+ \alpha^{A(\boldsymbol{\Omega}_A)}_{zz}(\omega)\eta^{B(\boldsymbol{\Omega}_B)}_{zz}(\omega)+A\leftrightarrow B\right] .
\end{equation}

From Eqs.~(\ref{etaFourierOmegaxx})-(\ref{alphaFourierOmega3}), we find
\begin{eqnarray}
    E_{\uparrow\uparrow} &=& -\frac{\hbar}{512\pi^3\varepsilon^2_0 R^6}\int_{-\infty}^{\infty} \!\!\! d\omega \left[ \alpha^{A}(\omega_{A+})\eta^{B}(\omega_{B+})+  \alpha^{A}(\omega_{A-})\eta^{B}(\omega_{B-})+9\alpha^{A}(\omega_{A+})\eta^{B}(\omega_{B-}) \right. \cr\cr
    &&
    + \left.  9\alpha^{A}(\omega_{A-})\eta^{B}(\omega_{B+})+4\alpha^{A}(\omega)\eta^{B}(\omega)+A\leftrightarrow B \right] \label{EUUalphaeta} .
\end{eqnarray}
Applying a change of variables -- similar to that used in the previous case --, we obtain
\begin{equation}
     E_{\uparrow \uparrow} = \mathcal{E}(\Omega_A-\Omega_B)+9\mathcal{E}(\Omega_A+\Omega_B)+2\mathcal{E}(0) \, . \label{smuu}
\end{equation}

\subsection{Spheres spinning perpendicular to each other with one of them parallel to the line joining them $(\uparrow\rightarrow)$}

In this configuration, shown in Fig.~\ref{Fig:configurations}(c), we take $\boldsymbol{\Omega}_A\parallel\boldsymbol{\hat{z}}$ and $\boldsymbol{\Omega}_B\parallel \boldsymbol{\hat{R}}\parallel \boldsymbol{\hat{x}}$.  Equations~(\ref{etaFourierOmegaxx})-(\ref{alphaFourierOmega3}) still apply to nanosphere $A$, but for nanosphere $B$, they must be changed to account for rotation around the $x$-axis. Implementing these substitutions, Eq.~(\ref{londongeral}) reduces to
\begin{equation}
    E_{\uparrow\rightarrow}= -\frac{\hbar}{128\pi^3\varepsilon^2_0 R^6}\int_{-\infty}^{\infty} \!\!\! d\omega \left[ 4\alpha^{A(\boldsymbol{\Omega}_A)}_{xx}(\omega)\eta^{B(\boldsymbol{\Omega}_B)}_{xx}(\omega)+\alpha^{A(\boldsymbol{\Omega}_A)}_{yy}(\omega)\eta^{B(\boldsymbol{\Omega}_B)}_{yy}(\omega)+ \alpha^{A(\boldsymbol{\Omega}_A)}_{zz}(\omega)\eta^{B(\boldsymbol{\Omega}_B)}_{zz}(\omega)+A\leftrightarrow B\right] , \label{eupright}
\end{equation}
which implies that
\begin{eqnarray}
    E_{\uparrow\rightarrow} &=& -\frac{\hbar}{512\pi^3\varepsilon^2_0 R^6}\int_{-\infty}^{\infty} \!\!\! d\omega \left\{ 8\left[\alpha^{A}(\omega_{A+})+\alpha^{A}(\omega_{A-})\right] \eta^{B}(\omega_{B})+2\alpha^{A}(\omega_{A})\left[\eta^{B}(\omega_{B+})+\eta^{B}(\omega_{B-})\right] \right. \cr\cr
    &&+ \left. \left[\alpha^{A}(\omega_{A+})+\alpha^{A}(\omega_{A-})\right]\left[\eta^{B}(\omega_{B+})+\eta^{B}(\omega_{B-})\right]
     \right\} +A\leftrightarrow B \label{EUUalphaeta} \,.
\end{eqnarray}
After performing an appropriate change of variables, we obtain
\begin{equation}
      E_{\uparrow\rightarrow } = 8\mathcal{E}(\Omega_A)+2\mathcal{E}(\Omega_B)+\mathcal{E}(\Omega_A-\Omega_B)+\mathcal{E}(\Omega_A+\Omega_B) \, . \label{smur}
\end{equation}

\subsection{Spheres spinning perpendicular to each other with both perpendicular to the line joining them $\uparrow\odot$}

Finally, we take $\boldsymbol{\Omega}_A\parallel\boldsymbol{\hat{z}}$, $\boldsymbol{\Omega}_B\parallel \boldsymbol{\hat{y}}$ and $\boldsymbol{\hat{R}}\parallel\boldsymbol{\hat{x}}$, as illustrated in Fig.~\ref{Fig:configurations}(d). In this case, Eq.~(\ref{eupright}) is still valid. Substituting the relations (\ref{etaFourierOmegaxx})-(\ref{alphaFourierOmega3}), adapted for the new rotation axis of nanosphere $B$, we obtain
\begin{eqnarray}
    E_{\uparrow\odot}&=& -\frac{\hbar}{256\pi^3\varepsilon^2_0 R^6}\int_{-\infty}^{\infty} \!\!\! d\omega \left\{ \left[\alpha^{A}(\omega_{A+})+\alpha^{A}(\omega_{A-})\right]\eta^{B}(\omega_{B})+\alpha^{A}(\omega_{A})\left[\eta^{B}(\omega_{B+})+\eta^{B}(\omega_{B-}) \right] \right. \cr\cr
    &&+ \left. \left[(\alpha^{A}(\omega_{A+})+\alpha^{A}(\omega_{A-})\right]\left[\eta^{B}(\omega_{B+})+\eta^{B}(\omega_{B-})\right]
     \right\}+A\leftrightarrow B \label{EUUalphaeta} \,,
\end{eqnarray}
which can be rewritten as
\begin{equation}
      E_{\uparrow\odot} = 2\mathcal{E}(\Omega_A)+2\mathcal{E}(\Omega_B)+4\mathcal{E}(\Omega_A-\Omega_B)+4\mathcal{E}(\Omega_A+\Omega_B) \, . \label{smuo}
\end{equation}

\section{Estimate of the total vdW force between nanospheres}

In this section, we estimate the total vdW force between the nanospheres at rest. 
Note the dielectric model 
for the polaritonic GHz resonance
described in the letter underestimates the interaction since it does not account for resonances at much higher frequencies. 
On the other hand, such high-frequency resonances provide a negligible rotational effect as the spinning frequencies cannot exceed the GHz range. 
Thus, we focus on their contribution to the static vdW interaction energy, which can 
 be obtained by setting $\Omega=0$ in any of the Eqs.~(\ref{smrr}), (\ref{smuu}), (\ref{smur}) or (\ref{smuo}). 
 After performing a Wick rotation, this energy can be written as a sum over the Matsubara frequencies $\xi_n=2\pi\,n\,k_BT/\hbar$ along the imaginary frequency axis ~\cite{IsraelachviliBook}: 
\begin{equation}\label{vdW_Wick}
    E_{\rm vdW}^{(0)}=-\frac{6k_BTa^6}{R^6}\sum_{n=0}^{\infty}{}'\frac{\varepsilon_A(i\xi_n)-\varepsilon_0}{\varepsilon_A(i\xi_n)+2\varepsilon_0} \,\frac{\varepsilon_B(i\xi_n)-\varepsilon_0}{\varepsilon_B(i\xi_n)+2\varepsilon_0} \, ,
\end{equation}
where 
the primed sum means that the $n=0$ term is multiplied by $1/2.$
As the permittivity of BST is not known in the IR and UV domains, we estimate the vdW interaction energy (\ref{vdW_Wick}) by comparing it with the expression for the Hamaker constant $H\equiv-12\pi\, D^2 \,U_{\rm vdW}(D).$  
Here,  $U_{\rm vdW}(D)$ denotes the non-retarded vdW interaction energy per unit area between two planar surfaces of BST separated by a distance $D.$
Starting from the Lifshitz formula and
taking the non-retarded 
and single round-trip approximations~\cite{Genet2004}, one finds~\cite{IsraelachviliBook}
\begin{equation} \label{H}
    H \approx \frac{3}{2}k_BT\,\sum_{n=0}^{\infty}{}'\left[\frac{\varepsilon_{\rm BST}(i\xi_n)-\varepsilon_0}{\varepsilon_{\rm BST}(i\xi_n)+\varepsilon_0}\right]^2 \, ,
\end{equation}
where $\varepsilon_{\rm BST}$ is the the permittivity of BST. 
We now take $\varepsilon_A=
\varepsilon_B=\varepsilon_{\rm BST}$ in the 
 right-hand-side of 
(\ref{vdW_Wick}) and approximate the resulting Matsubara sum so as to connect it with (\ref{H}): $ E_{\rm vdW}^{(0)}\approx -(16/9)\,H\, (a/R)^6.$
Most dielectric materials have Hamaker constants close to $H\sim 5\times 10^{-20}\,{\rm J}$ \cite{Hough1980}. Taking the geometric parameters $a=60\,{\rm nm}$ and 
$R=180\,{\rm nm},$ we estimate the magnitude of the (attractive) vdW force between the BST nanospheres as $ |F_{\rm vdW}^{(0)}|\sim 4 \,{\rm fN}.$

\section{Interaction energy obtained from a naive application of FDT in the rotating frame}

We can quantify the nonequilibrium effects by comparing our results with those that would be obtained under the assumption of thermal equilibrium. This can be done by employing the polarizability tensor given in Eqs.~(\ref{alphaFourierOmega1})-(\ref{alphaFourierOmega3}), but replacing the Hadamard function with the FDT relation (\ref{FDT}), instead of using Eqs.~(\ref{etaFourierOmegaxx})-(\ref{etaFourierOmega}). In this case, the interaction energy can no longer be cast in terms of the auxiliary function $\mathcal{E}(\omega)$, reflecting the incompatibility between the FDT and the rotational Doppler effect.

Here, we restrict our analysis to the $\rightarrow\rightarrow$ configuration, in which $\boldsymbol{\Omega}_\zeta=\Omega_\zeta \boldsymbol{\hat{z}}$ and $\boldsymbol{R}=R\boldsymbol{\hat{z}}$. However, similar considerations apply to other configurations. The interaction energy becomes
\begin{equation}
    E^{T=0}_{\rightarrow\rightarrow} = -\frac{\hbar}{64\pi^3\varepsilon^2_0 R^6}\int_{-\infty}^{\infty} d\omega \left[ \alpha^{A(\boldsymbol{\Omega}_A)}_{xx}(\omega)\eta^{B(\boldsymbol{\Omega}_B, T=0)}_{xx}(\omega)+ 2\alpha^{A(\boldsymbol{\Omega}_A)}_{zz}(\omega)\eta^{B(\boldsymbol{\Omega}_B,T=0)}_{zz}(\omega)+A\leftrightarrow B\right] , \label{errthermaleqpartial}
\end{equation}
where we used that $\alpha^{A(\boldsymbol{\Omega}_A)}_{xx}(\omega)=\alpha^{A(\boldsymbol{\Omega}_A)}_{yy}(\omega)$, $\eta^{B(\boldsymbol{\Omega}_B, T=0)}_{xx}(\omega)=\eta^{B(\boldsymbol{\Omega}_B, T=0)}_{yy}(\omega)$, and that the off-diagonal elements of the Hadamard function vanish in thermal equilibrium. This is a key distinction from the nonequilibrium case [Eq.~(\ref{errnonequilibriumgeneral})], where the presence of nondiagonal terms led to an interaction energy that depended solely on the relative angular velocity -- a feature that no longer holds here. 

To proceed, we analyze each contribution separately. Defining $\Lambda_{xx}\equiv \int_{-\infty}^{\infty} d\omega [\alpha^A_{xx}(\omega)\eta^B_{xx}(\omega) +\eta^A_{xx}(\omega)\alpha^B_{zz}(\omega)]$, we see from Eq.~(\ref{FDT}) that
\begin{equation}
\Lambda_{xx} = 2\int_{-\infty}^{\infty} d\omega \left\{ \alpha^{A(\boldsymbol{\Omega}_A)}_{xx}(\omega)\,\mbox{sgn}(\omega)\,\mbox{Im}\left[ \alpha_{xx}^{B(\boldsymbol{\Omega}_B)}(\omega) \right] +\,\mbox{sgn}(\omega)\,\mbox{Im}\left[ \alpha_{xx}^{A(\boldsymbol{\Omega}_A)}(\omega) \right] \alpha^{B(\boldsymbol{\Omega}_B)}_{xx}(\omega) \right\} \, .
\end{equation}
Recalling that the real (imaginary) part of the polarizability is an even (odd) function of $\omega$, we find 
\begin{equation}
    \Lambda_{xx} = 4\,\mbox{Im}\int_{0}^{\infty}d\omega\,\alpha_{xx}^{A(\boldsymbol{\Omega}_A)}(\omega)\alpha_{xx}^{B(\boldsymbol{\Omega}_B)}(\omega) \, .
\end{equation}
Substituting this back into Eq.~(\ref{errthermaleqpartial}) -- with an analogous treatment for the $zz$ terms -- yields
\begin{equation}
     E^{T=0}_{\rightarrow\rightarrow}  = -\frac{\hbar}{16\pi^3\varepsilon^2_0 R^6} \,\mbox{Im}\int_{0}^{\infty}d\omega \left[\alpha_{xx}^{A(\boldsymbol{\Omega}_A)}(\omega)\alpha_{xx}^{B(\boldsymbol{\Omega}_B)}(\omega)+2\alpha_{zz}^{A(\boldsymbol{\Omega}_A)}(\omega)\alpha_{zz}^{B(\boldsymbol{\Omega}_B)}(\omega) \right]\label{londonani} \, .
\end{equation}
Using Eqs.~(\ref{alphaFourierOmega1})-(\ref{alphaFourierOmega3}), we obtain 
\begin{equation}
   E^{T=0}_{\rightarrow\rightarrow} = -\frac{\hbar}{16\pi^3\varepsilon_0^2R^6}\,\mbox{Im}\int_{0}^{\infty}d\omega \frac{\left[\alpha^{A}(\omega_{A+})+\alpha^{A}(\omega_{A-}) \right]\left[ \alpha^{B}(\omega_{B+})+\alpha^{B}(\omega_{B-})\right] + 8\alpha^{A}(\omega) \alpha^{B}(\omega)}{4} \, . \label{upup}
\end{equation}
We see that, unlike in the full nonequilibrium treatment, the integrand cannot be written only in terms of the relative angular velocity. This contradicts the symmetry of the configuration and highlights the inadequacy of assuming thermal equilibrium when the nanospheres are spinning.

\end{widetext}












\end{document}